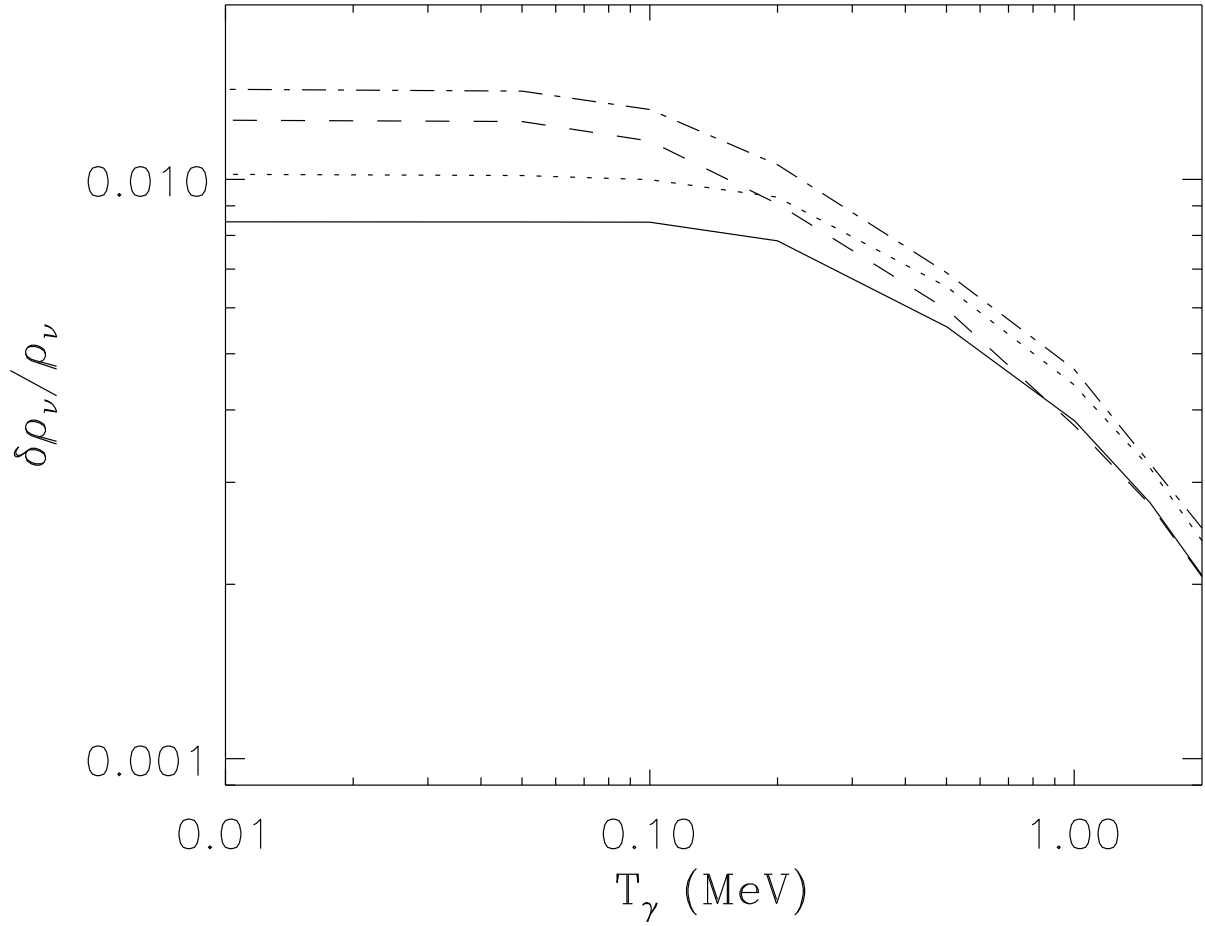

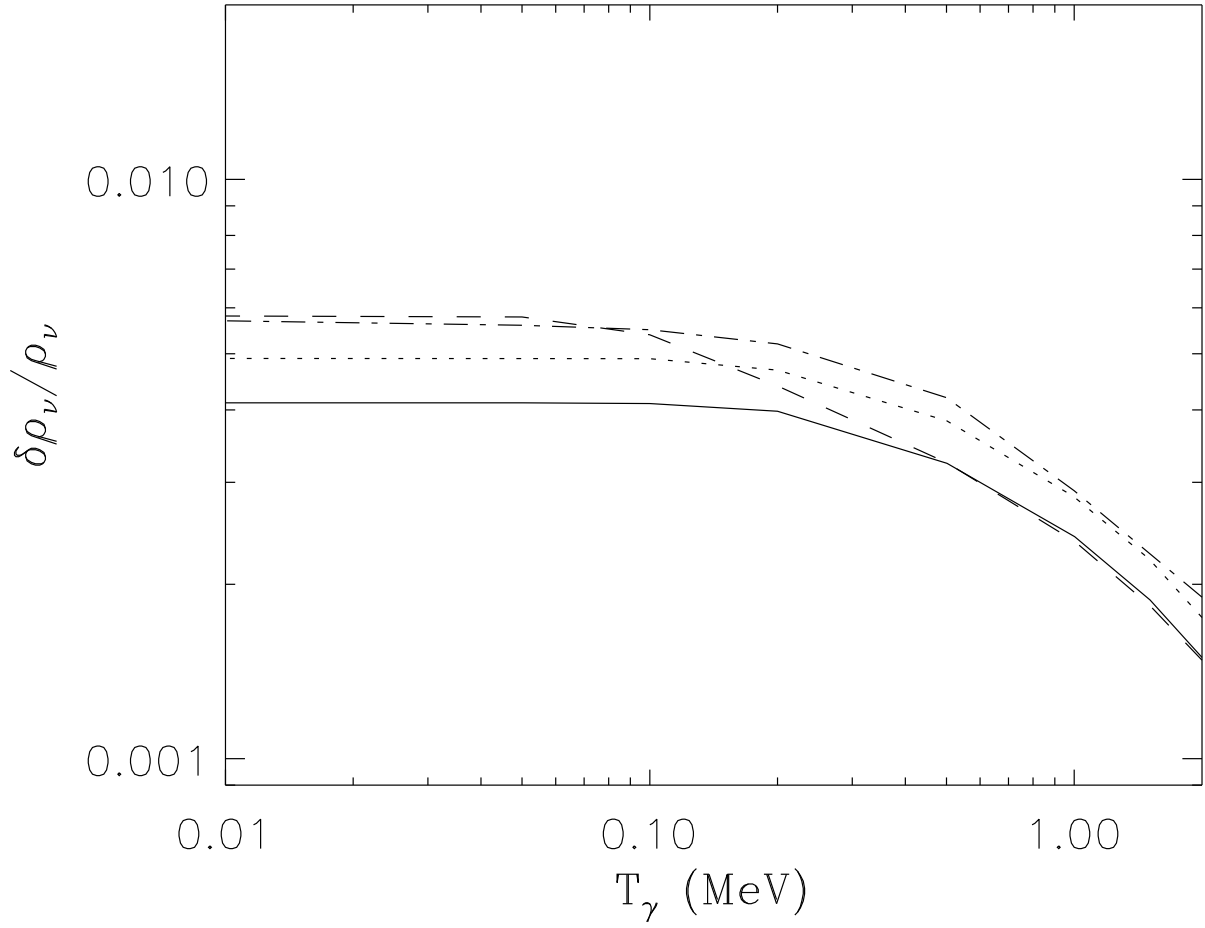

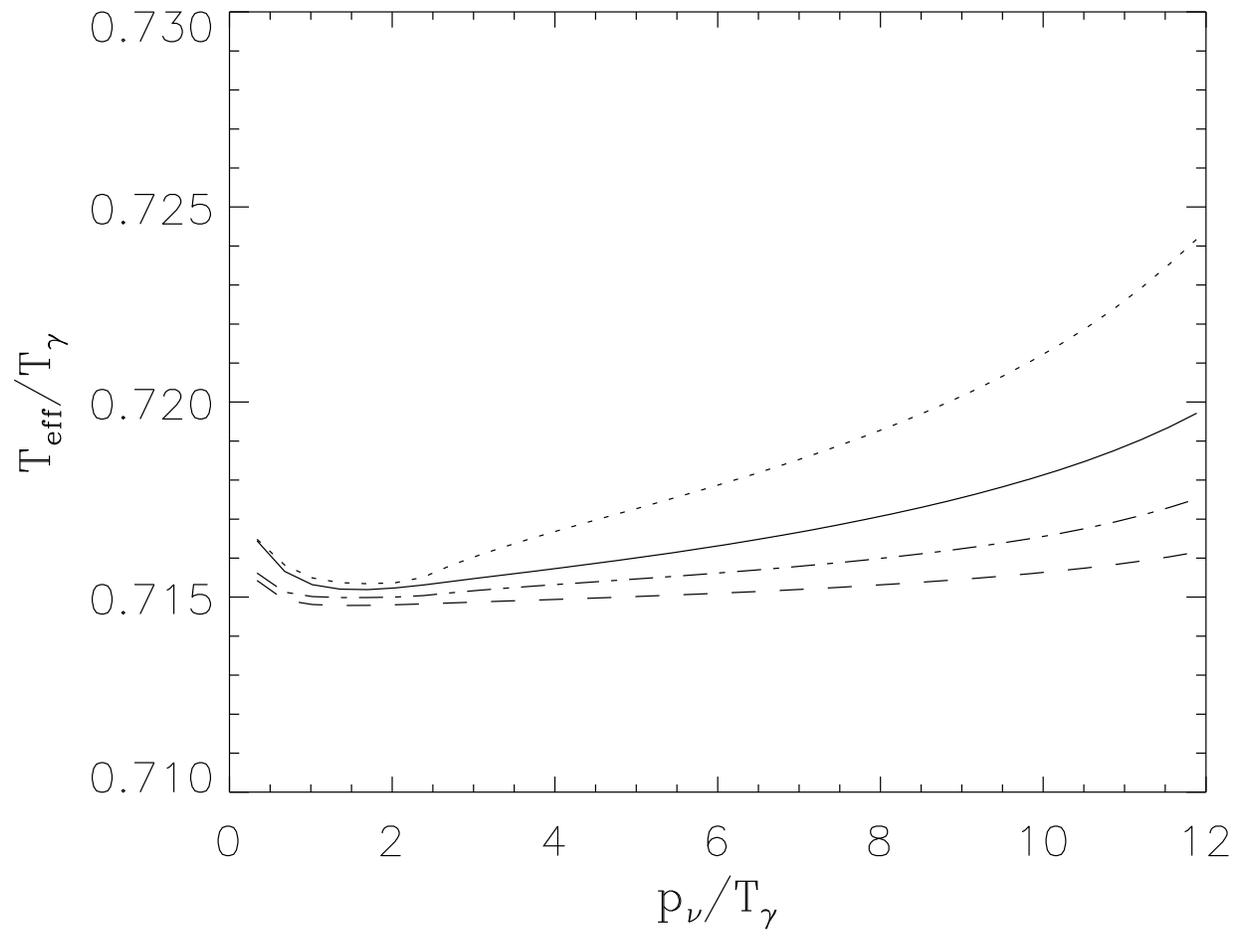

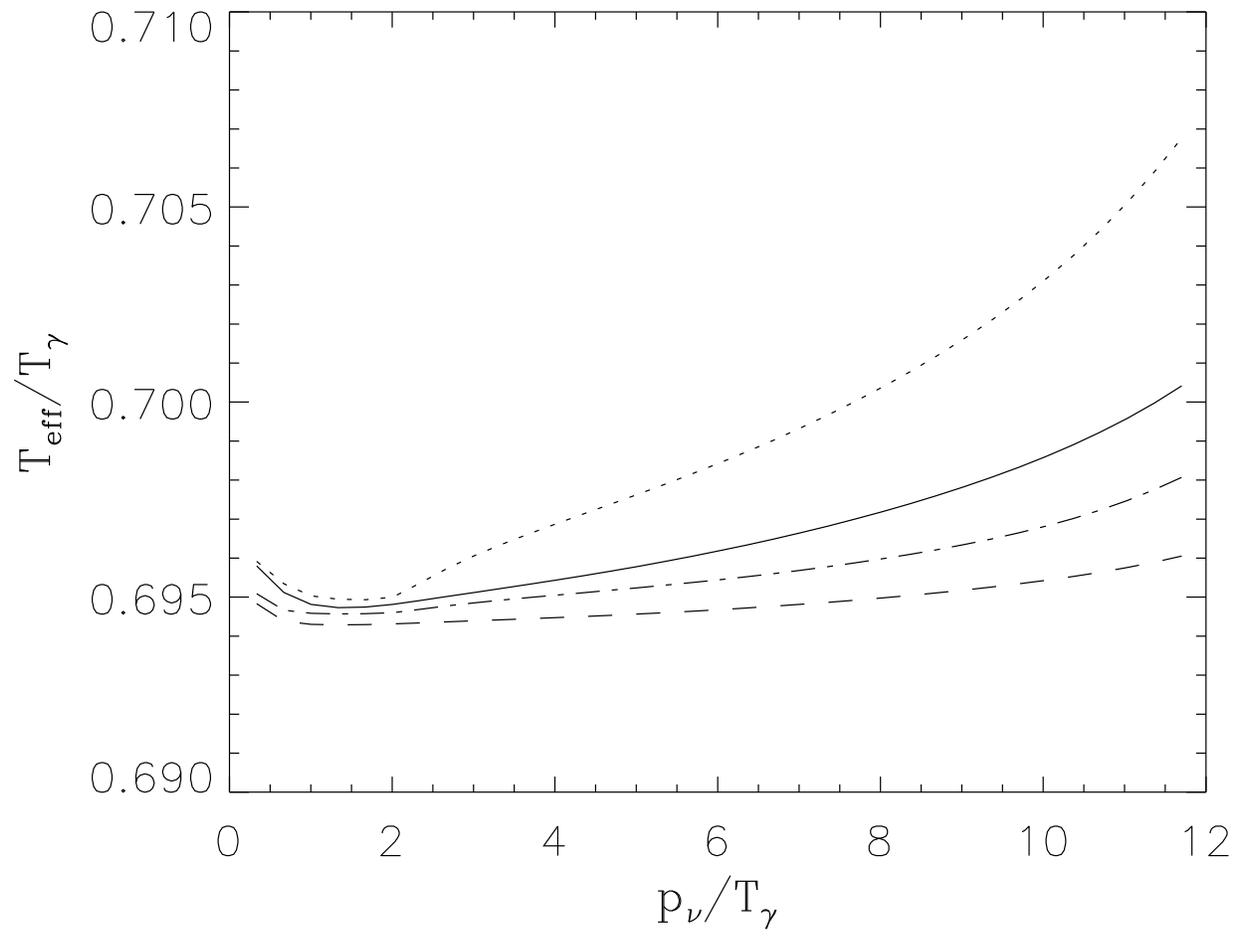

# Neutrino decoupling in the early Universe

Steen Hannestad and Jes Madsen

*Institute of Physics and Astronomy, University of Aarhus, DK-8000 Århus C, Denmark*

(April 4, 1995)

## Abstract

A calculation of neutrino decoupling in the early Universe, including full Fermi-Dirac statistics and electron mass dependence in the weak reaction rates, is presented. We find that after decoupling, the electron neutrinos contribute 0.83% more to the relativistic energy density than in the standard scenario, where neutrinos are assumed not to share the heating from $e^{\pm}$ annihilation. The corresponding number for muon and tau neutrinos is 0.41%. This has the consequence of modifying the primordial $^4$He abundance by $\Delta Y = +1.0 \times 10^{-4}$, and the cosmological mass limit on light neutrinos by 0.2–0.5 eV.

95.30.Cq, 13.15.+g, 14.60.Pq, 98.80.Ft

Typeset using REVTEX

# I. INTRODUCTION

The phenomenon of non-equilibrium thermodynamics is of fundamental interest to the understanding of the early evolution of the Universe. One example of this non-equilibrium behaviour is the decoupling of neutrinos from the electromagnetic plasma, at a temperature of a few MeV. Usually, it is assumed that the neutrinos decouple before the temperature gets below the mass of electrons and positrons. Thus they do not share the entropy transfer from e$^{\pm}$ to photons. This means that, although the neutrino distribution is still described by an equilibrium distribution after decoupling [1], the temperature is only $(4/11)^{1/3} \approx 0.714$ times the photon temperature.

However, the fact that the decoupling temperature is of the same order as the electron mass, means that neutrinos must, to a small degree, share the entropy transfer, leading to what is known as neutrino heating. This observation has led to several re-investigations of neutrino decoupling [2–6]. All of these find a change in the final energy density of neutrinos relative to the standard case of total decoupling of around 1%. One problem is, however, that they all assume that Boltzmann statistics is adequate for treating this problem. Furthermore they assume, that the electron mass is negligible in the weak interaction matrix elements. An approximation that is definitely valid whenever the temperature is very large, but possibly not when it is comparable to $m_e$. It is therefore of relevance to investigate whether the results obtained by using these approximations are correct.

Since the neutrino distributions are important both in Big Bang nucleosynthesis calculations and structure formation models, it is of considerable interest to see if neutrino heating has any consequences for either.

# II. FUNDAMENTAL EQUATIONS

The fundamental equation governing the evolution of all particle abundances is the Boltzmann equation

$$C_\mathrm{L}[f] = \sum C_\mathrm{coll}[f], \tag{1}$$

where $C_\mathrm{L}$ is the Liouville operator and $\sum C_\mathrm{coll}$ is the sum of all possible collisional interactions. In the following we will assume that all distributions are homogeneous and isotropic, that is $f(\mathbf{x}, \mathbf{p}, t) = f(p, t)$. In that case, the Liouville operator is [7]

$$C_\mathrm{L}[f] = \frac{\partial f}{\partial t} - \frac{\mathrm{d}R}{\mathrm{d}t} \frac{1}{R} p \frac{\partial f}{\partial p}. \tag{2}$$

In the case of standard massless Dirac neutrinos, there are several contributions to $C_\mathrm{coll}$ [2,8]. To first order in the weak coupling constant all interactions are 2 particle scatterings and annihilations. All of these reactions and the corresponding matrix elements have been summarized in Tables I and II. Higher order contributions as well as electromagnetic corrections [9] to these lowest order terms are neglected in the following. Since there are only 2-particle interactions like $1 + 2 \to 3 + 4$, $C_\mathrm{coll}$ can be written as

$$C_\mathrm{coll}[f] = \frac{1}{2E_1} \int d^3\tilde{p}_2 d^3\tilde{p}_3 d^3\tilde{p}_4 \Lambda(f_1, f_2, f_3, f_4) \times \tag{3}$$
$$S \, |M|^2_{12 \to 34} \, \delta^4(p_1 + p_2 - p_3 - p_4)(2\pi)^4,$$

where $\Lambda(f_1, f_2, f_3, f_4) = (1 - f_1)(1 - f_2)f_3 f_4 - (1 - f_3)(1 - f_4)f_1 f_2$ is the phase space factor, including Pauli blocking of the final states, and $d^3\tilde{p} = d^3p/((2\pi)^3 2E)$. $S$ is a symmetrization factor of $1/2!$ for each pair of identical particles in initial or final states [10], and $|M|^2$ is the weak interaction matrix element squared, summed over the spin states of all particles except the one under scrutiny. $p_i$ is the four-momentum of particle $i$. Note that the before mentioned assumptions of Boltzmann statistics and zero electron mass simplify the collision integral tremendously, but it is in fact possible to evaluate most of it analytically without them.

Assuming standard weak interactions as described in Tables I and II, it is possible to analytically integrate this 9 dimensional integral down to two dimensions (Appendix). This makes the integral much easier to evaluate numerically.

The l.h.s. of the Boltzmann equation contains the factor $R^{-1}\mathrm{d}R/\mathrm{d}t$, which cannot be immediately evaluated. To do this we have to consider entropy conservation. Let $g_\star$ be the effective number of relativistic degrees of freedom [1]. If $g_*$ is constant in time we obtain the standard result of $R(t) \propto t^{\frac{1}{2}}$, but if this is not the case (e.g. during $e^{\pm}$ annihilation), then this is no longer true. Instead we write $R(t) \propto t^{\frac{1}{2}} f(t)$, where $f(t)$ can be found using the fact that entropy is conserved in the early Universe. In the case of totally decoupled neutrinos, $f(t)$ can be calculated without too much trouble. We have used this value for $f(t)$ in our calculations as the error induced is small [11]. The photon temperature is calculated at each time step by assuming $f^{-1} df/dt \ll \frac{1}{2} t^{-1}$, which is, at all times, an approximation better than $O(10^{-3})$. Then the Friedmann equation is

$$\left( \frac{\mathrm{d}R}{\mathrm{d}t} \frac{1}{R} \right)^2 = \frac{1}{4} t^{-2} = \frac{8\pi G \rho}{3}, \tag{4}$$

and knowing $\rho, \rho_\nu$ we can calculate $\rho_\gamma$ by assuming that $e^{\pm}$ are kept in complete thermodynamic equilibrium by electromagnetic reactions with the photon gas. We are thus able to follow all relevant thermodynamic quantities as functions of time.

### III. NUMERICAL RESULTS

To compare with previous results [2,3], we have solved the Boltzmann equation numerically for several different cases: 1. $m_e \neq 0$ and FD statistics; 2. $m_e = 0$ and FD statistics; 3. $m_e \neq 0$ and MB statistics; 4. $m_e = 0$ and MB statistics. In all cases we assume zero chemical potentials for the particles involved.

Case 4 is the approximation used in previous studies. Case 1 is the correct approach. The other cases test the errors introduced by the different approximations. For each of these scenarios, the energy density deviation, defined as $\delta \rho_\nu / \rho_\nu = (\rho_\nu - \rho_{\nu_0}) / \rho_\nu$, has been calculated as a function of the photon temperature, $T_\gamma$. $\rho_{\nu_0}$ is the energy density in a neutrino species that has decoupled long before $e^{\pm}$ annihilation. For electron neutrinos, the result is shown in Fig. 1, whereas Fig. 2 shows the result for muon and tau neutrinos. In the limit of MB statistics and zero electron mass (case 4), we reproduce the results of Ref. [2] to within 10-15%. This is quite reassuring, as we use a completely different procedure for solving the Boltzmann equations (Dodelson and Turner [2] use a first order perturbation expansion). It is seen that, using the correct FD statistics and electron mass (case 1), the end result of $\delta\rho_\nu/\rho_\nu$ is around 0.83% for $\nu_e$ and 0.41% for $\nu_{\mu,\tau}$. Generally we see that the deviation is smaller if FD statistics is used. This is because the phase space factors, $\Lambda$, in the collision integral get smaller. Using the correct treatment of electron mass also lessens the deviation, due to the changes in matrix element structure.

Fig. 3 shows the effective neutrino temperature, defined as [2]

$$T_{\text{eff}}(p) \equiv \frac{p}{\log(1/f(p) - 1)}, \tag{5}$$

for FD statistics with and without electron mass. Fig. 4 shows the same, but for MB statistics ($T_{\text{eff}} \equiv -p/\log f$). The reason for the offset between the two types of statistics is that, for FD statistics, the final neutrino temperature for totally decoupled neutrinos is $(4/11)^{1/3} \simeq 0.714$ [1], whereas for MB statistics it is $(1/3)^{1/3} \simeq 0.693$. We see that, regardless of the statistics used, the effective temperature rises with momentum for medium and high momentum states. As noted in Refs. [2,3], this is not surprising, because the weak cross sections are much larger for large momenta. The shape of the effective temperature curve is the same in all four cases, but the actual numerical values are slightly different, being higher if the electron mass is neglected in $C_{\text{coll}}[f]$. This effect reflects that the neutrino energy density is slightly higher for massless electrons, as seen in Fig. 1. For the lowest momentum states the effective temperature rises again. This may seem surprising, but it has to do with the integrated matrix elements (the functions called $F(p_1, p_2, p_3)$ in the Appendix). These functions contain factors that grow large for very small momenta, leading to a stronger interaction of very low momentum states, both with e$^\pm$ and with high momentum neutrinos. This means that the lowest momentum states are actually kept closer to thermal equilibrium than the medium momentum states.

# IV. DISCUSSION AND CONCLUSION

We have calculated neutrino decoupling in the early Universe by solving the Boltzmann equations. Our method differs from previous studies in that it makes no approximations for the distribution functions and the electron mass. Although our results do not deviate dramatically from these previous calculations, we show that the inclusion of FD statistics and non-zero electron mass reduce the effect of neutrino heating on the relativistic neutrino energy density by almost 50% for $\nu_e$ compared to the results of Dodelson & Turner [2].

We find that the neutrino distribution after decoupling is non-thermal at the 1% level. This is also in relatively good agreement with previous results [2,3]. The effective temperature grows with momentum for medium and large momenta, because of the momentum dependence of the weak interactions. The very low momentum states, however, interact more strongly, leading to a rise in the effective temperature for very small values of $p/T$.

The slight heating of neutrinos relative to the standard scenario has consequences for Big Bang nucleosynthesis. We have changed the nucleosynthesis code of Kawano [12] to include the effect of neutrino heating. As discussed in Ref. [13], there are several effects that combine to change the nucleosynthesis scenario. First of all, the energy density in neutrinos is changed. However, this has the consequence of lowering the energy density in photons and $e^\pm$, because of energy conservation. The result is, if we use FD statistics and non-zero electron mass, that the photon temperature goes up because of $e^\pm$ annihilations by a factor of 1.3998 instead of the usual 1.4010 (a change of $-0.09\%$). Furthermore, the weak rates are changed, because of the different abundances of $e^\pm$ and $\nu_e$. With the relevant changes to the code, we obtain a change in the primordial $^4$He abundance, $Y$, of $\Delta Y = +1.0 \times 10^{-4}$. This is in good agreement with Ref. [13]. They calculate a change of $+1.5 \times 10^{-4}$, but with a 50% larger neutrino heating. Other authors have found similar values [3,4]. A change in $^4$He abundance of $= +1.0 \times 10^{-4}$ is much below current observational accuracies. The systematic uncertainty in the primordial $^4$He abundance is estimated to be as large as $\Delta Y_{\text{sys}} = \pm 0.015$ [14], or more than a factor of 100 larger than the change induced by neutrino heating.

Another consequence of neutrino heating is that it increases the number density of neutrinos. If one of the neutrino species has a mass, it will therefore contribute slightly more than usually expected to $\Omega$, the cosmological density parameter. In general, it is possible to put a mass limit on a light neutrino that is completely decoupled long before $e^{\pm}$ annihilation [15]. The mass density of such a species today is

$$\rho_\nu = n_\nu m_\nu. \tag{6}$$

Expressing this in terms of the photon density, $n_\gamma = 2\zeta(3)T_\gamma^3/\pi^2$, one gets

$$\rho_\nu = m_\nu \frac{n_\nu}{n_\gamma} \zeta(3) \frac{2}{\pi^2} T_\gamma^3. \tag{7}$$

If the neutrinos decouple long before $e^{\pm}$ annihilation, this can be translated into the normal textbook relation

$$\Omega_\nu h^2 = \frac{g_\nu}{2} \frac{m_\nu}{93.03 \text{eV}}, \tag{8}$$

using a present photon temperature of 2.736 K. $h$ is the dimensionless Hubble constant, and $g_\nu = 2$ for one flavor of neutrino and antineutrino. Since observations demand that $\Omega_\nu \leq 1$, we have a mass limit on any given light neutrino. Because of neutrino heating, this limit is changed by a small amount. Using FD statistics and non-zero $m_e$, the final number density of neutrinos after decoupling deviates from the standard case by $\delta n_\nu/n_\nu \sim 0.52\%$ for $\nu_e$ and $\delta n_\nu/n_\nu \sim 0.25\%$ for $\nu_{\mu,\tau}$. For electron neutrinos this changes Eq. (8) to

$$\Omega_{\nu_e} h^2 = \frac{g_{\nu_e}}{2} \frac{m_{\nu_e}}{92.55 \text{eV}}, \tag{9}$$

whereas for muon or tau neutrinos we find

$$\Omega_{\nu_{\mu,\tau}} h^2 = \frac{g_{\nu_{\mu,\tau}}}{2} \frac{m_{\nu_{\mu,\tau}}}{92.80 \text{eV}}. \tag{10}$$

As this is a very small change to the standard value, neutrino heating does not alter the usual conclusion, that electron neutrinos cannot contribute more than about 0.25 (using the current experimental upper limit to its mass, $m_{\nu_e} \leq 7$ eV [16]) to $\Omega$, whereas muon and tau neutrinos can.

Thus, it is safe to ignore neutrino heating when doing nucleosynthesis and structure formation calculations.

## ACKNOWLEDGMENTS

This work was supported in part by the Theoretical Astrophysics Center under the Danish National Research Foundation, and by the Theoretical Astroparticle Network under the European Human Capital and Mobility Program.

## APPENDIX: COLLISION INTEGRALS

This appendix shows how to reduce the integral in Eq. (3) from 9 to 2 dimensions. Except for the inclusion of mass terms, most of what can be found in this appendix was originally developed by Yueh & Buchler [17] for use in supernova calculations.

The collision integrals all have the form of Eq. (3). The innermost integral can be rewritten using the equality

$$\frac{d^3 p_4}{2E_4} = d^4 p_4 \delta(p_4^2 - m_4^2) \Theta(p_4^0). \tag{A1}$$

The integral over $d^4 p_4$ is now done using the $\delta$-function. Hereafter $p_4$ is fixed as

$$p_4^2 = p_1^2 + p_2^2 + p_3^2 + 2(p_1 \cdot p_2 - p_1 \cdot p_3 - p_2 \cdot p_3). \tag{A2}$$

Now, the following angles are introduced

$$\cos \alpha = \frac{\mathbf{p_1} \cdot \mathbf{p_2}}{p_1 p_2} \tag{A3}$$

$$\cos \theta = \frac{\mathbf{p_1} \cdot \mathbf{p_3}}{p_1 p_3} \tag{A4}$$

$$\cos \alpha' = \frac{\mathbf{p_2} \cdot \mathbf{p_3}}{p_2 p_3} \tag{A5}$$

$$= \cos \alpha \cos \theta + \sin \alpha \sin \theta \cos \beta.$$

Thus we obtain

$$d^3p_2 = p_2^2 dp_2 d\cos\alpha d\beta \tag{A6}$$

$$d^3p_3 = p_3^2 dp_3 d\cos\theta d\mu. \tag{A7}$$

The integration over $d\beta$ is carried out using the $\delta$-function. We use that

$$p_4^2 - m_4^2 = f(\beta). \tag{A8}$$

An important relation for $\delta$-functions is

$$\int d\beta \delta(f(\beta)) = \sum_i \int d\beta \frac{1}{\left|\frac{df(\beta)}{d\beta}\right|_{\beta=\beta_i}} \delta(\beta - \beta_i), \tag{A9}$$

where $\beta_i$ are the roots of $f(\beta) = 0$. The derivative is evaluated to

$$\frac{df(\beta)}{d\beta} = -2p_2 p_3 \sin\alpha \sin\theta \sin\beta. \tag{A10}$$

$\sin\beta_i$ is found as $\pm(1 - \cos^2\beta_i)^{1/2}$, where

$$\begin{aligned}
\cos\beta_i = &\frac{2E_2 E_3 - 2p_2 p_3 \cos\alpha\cos\theta - Q}{2p_2 p_3 \sin\alpha\sin\theta} \\
&- \frac{(2E_1 E_2 - 2p_1 p_2 \cos\alpha)}{2p_2 p_3 \sin\alpha\sin\theta} \\
&+ \frac{(2E_1 E_3 - 2p_1 p_3 \cos\theta)}{2p_2 p_3 \sin\alpha\sin\theta},
\end{aligned} \tag{A11}$$

and we have introduced $Q = m_1^2 + m_2^2 + m_3^2 - m_4^2$. The equation for $\sin\beta_i$ will have two solutions, one in the interval $[0, \pi]$, and one in the interval $[\pi, 2\pi]$. Since the r.h.s. of Eq. (A9) is symmetric in $\beta$ we can multiply by two and integrate over $[0, \pi]$. The limits on the integral $d\cos\alpha$ come from demanding that $\cos^2(\beta) \leq 1$. But this means that

$$(2p_2 p_3 \sin\alpha\sin\theta\sin\beta)^2 \geq 0. \tag{A12}$$

Notice, that this is the same as demanding that

$$\left(\frac{df(\beta)}{d\beta}\right)^2 \geq 0. \tag{A13}$$

Finally we end up with

$$\int_0^{2\pi} d\beta \, \delta(f(\beta)) = 2 \frac{1}{\left| \frac{df(\beta)}{d\beta} \right|_{\beta=\beta_i}} \Theta \left( \left| \frac{df(\beta)}{d\beta} \right|_{\beta=\beta_i}^2 \right). \tag{A14}$$

The derivative can be written as

$$\left| \frac{df(\beta)}{d\beta} \right|_{\beta=\beta_i} = \sqrt{a \cos^2 \alpha + b \cos \alpha + c}, \tag{A15}$$

where

$$a = p_2^2(-4\kappa + 8\epsilon) \tag{A16}$$

$$b = p_2(p_1 - \epsilon/p_1)(8\gamma + 4Q + 8\epsilon) \tag{A17}$$

$$c = -4\gamma^2 - 4\gamma Q - Q^2 - 8\gamma\epsilon - 4Q\epsilon - 4\epsilon^2 \tag{A18}$$

$$+4p_2^2 p_3^2 (1 - \cos^2 \theta),$$

and we have introduced the following parameters in order to limit the amount of space required to write out the formulae

$$\gamma = E_1 E_2 - E_1 E_3 - E_2 E_3 \tag{A19}$$

$$\epsilon = p_1 p_3 \cos \theta \tag{A20}$$

$$\kappa = p_1^2 + p_3^2. \tag{A21}$$

Notice that Eqs. (A16), (A17) and (A18) reduce to those of Yueh & Buchler [17] in the limit of zero masses.

Now, any one of the possible matrix elements only include products of 4-momenta. All the possible products of these momenta are calculated below:

$$p_1 \cdot p_2 = E_1 E_2 - p_1 p_2 \cos \alpha \tag{A22}$$

$$p_1 \cdot p_3 = E_1 E_3 - p_1 p_3 \cos \theta \tag{A23}$$

$$p_1 \cdot p_4 = m_1^2 + (E_1 E_2 - p_1 p_2 \cos \alpha) \tag{A24}$$

$$-(E_1 E_3 - p_1 p_3 \cos \theta)$$

$$p_2 \cdot p_3 = (E_1 E_2 - p_1 p_2 \cos \alpha) \tag{A25}$$

$$-(E_1 E_3 - p_1 p_3 \cos \theta) + Q/2$$

$$p_2 \cdot p_4 = (E_1 E_3 - p_1 p_3 \cos \theta) + m_2^2 - Q/2 \qquad (A26)$$

$$p_3 \cdot p_4 = (E_1 E_2 - p_1 p_2 \cos \alpha) - m_3^2 + Q/2. \qquad (A27)$$

Because all the above products are analytically integrable over $d \cos \alpha$, the integrals over this parameter can now be carried out by use of the fundamental relation

$$\int_{-\infty}^{\infty} \frac{dx}{\sqrt{ax^2 + bx + c}} \Theta(ax^2 + bx + c) = \frac{\pi}{\sqrt{-a}} \Theta(b^2 - 4ac) \qquad (A28)$$

The step function comes from demanding that there be a real integration interval. This actually also insures that the roots of $ax^2 + bx + c$ are not outside the fundamental integration interval of $[-1, 1]$, because the step function singles out the physical situations, where $d \cos \alpha$ cannot be outside this interval. Integrating over $d\mu$ is no problem, as there is no dependency on this parameter. The final integration that can be done is the one over $d \cos \theta$. Any one of the possible products of momenta is analytically integrable over $d \cos \theta$. The solution of $b^2 - 4ac = 0$ gives the integration interval. The solutions are

$$\cos \theta = (-2\gamma - 2p_2^2 - Q \qquad (A29)$$
$$\pm 2p_2(2\gamma + p_1^2 + p_2^2 + p_3^2 + Q)^{\frac{1}{2}})/(2p_1 p_3).$$

If there is to be a real integration interval, both of these solutions must be real. The fundamental integration interval is of course $[-1, 1]$, but the real limits are $\alpha = \sup[-1, \cos \theta_{\min}]$ and $\beta = \inf[1, \cos \theta_{\max}]$. There can only be a real integration interval if both $\alpha$ and $\beta$ are real numbers and $\beta \geq \alpha$.

Note that there are seemingly two places where divergences occur in the collision integral. The first one is for $p_1 = p_3, \cos \theta = 1$, but in this case there is no problem because although the integrand becomes infinite, the integral is finite [17]. The second place is for $p_1 \to 0$. To see what happens here, we have to go back to the fundamental integral, Eq. (A28). If $p_1 \to 0$ then $a = -4p_2^2 p_3^2$, but then there can be no divergence, as $p_1$ no longer appears in any denominator. Clearly there is still the possibility of divergence if $m_1 \to 0$ also, because of the $1/E_1$ term. Fortunately it turns out that $b^2 - 4ac = 0$ if $m_1, p_1 = 0$, so that the rate becomes equal to 0 in this case, as it ought to be. Finally the collision integral can be written as

$$C_{\text{coll}}[f] = \frac{2}{(2\pi)^4} \frac{1}{2E_1} \int \frac{p_2^2 dp_2}{2E_2} \frac{p_3^2 dp_3}{2E_3} S \times \tag{A30}$$
$$\Lambda(f_1, f_2, f_3, f_4) F(p_1, p_2, p_3) \Theta(A),$$

where $A$ is the parameter space allowed (that is, the space defined by requiring $\alpha, \beta$ real and $\beta \geq \alpha$). $F$ is the matrix element integrated over $d\cos\alpha$ and $d\cos\theta$,

$$F(p_1, p_2, p_3) = \int |M|^2 \, d\cos\alpha \, d\cos\theta \tag{A31}$$

FIGURES

FIG. 1. The evolution of $\delta\rho_\nu/\rho_\nu$ for electron neutrinos. The solid curve corresponds to case 1, the dashed to case 2, the dotted to case 3, and the dot-dashed to case 4.

FIG. 2. The evolution of $\delta\rho_\nu/\rho_\nu$ for muon and tau neutrinos. The curve labels are as in Fig. 1.

FIG. 3. The effective neutrino temperature after complete $e^\pm$ annihilation, using FD statistics. The solid line is for $\nu_e$ and case 1, the dashed is for $\nu_{\mu,\tau}$ also case 1. The dotted is for $\nu_e$ and case 2, the dot-dashed for $\nu_{\mu,\tau}$ and case 2.

FIG. 4. The effective neutrino temperature after complete $e^\pm$ annihilation, using MB statistics. The solid line is for $\nu_e$ and case 3, the dashed is for $\nu_{\mu,\tau}$ also case 3. The dotted is for $\nu_e$ and case 4, the dot-dashed for $\nu_{\mu,\tau}$ and case 4.

TABLE I. Possible electron neutrino processes and the corresponding matrix elements. 1 is defined equal to the particle for which we calculate $C_{\text{coll}}$. 2 is the other incoming particle. 3 is defined as either particle 1 going out (scattering) or the outgoing lepton (annihilation). 4 is defined as either particle 2 going out (scattering) or the outgoing antilepton (annihilation). We have introduced the quantities $C_V = \frac{1}{2} + 2\sin^2\theta_W$ and $C_A = \frac{1}{2}$, where $\sin^2\theta_W \approx 0.23$. $p_i$ is the four-momentum of particle $i$.

| Process | $S\,\lvert M\rvert^2$ |
|---|---|
| $\nu_e + \overline{\nu}_e \to \nu_e + \overline{\nu}_e$ | $128 G_F^2 (p_1 \cdot p_4)(p_2 \cdot p_3)$ |
| $\nu_e + \nu_e \to \nu_e + \nu_e$ | $32 G_F^2 (p_1 \cdot p_2)(p_3 \cdot p_4)$ |
| $\nu_e + \nu_i \to \nu_e + \nu_i$ | $32 G_F^2 (p_1 \cdot p_2)(p_3 \cdot p_4)$ |
| $\nu_e + \overline{\nu}_i \to \nu_e + \overline{\nu}_i$ | $32 G_F^2 (p_1 \cdot p_4)(p_2 \cdot p_3)$ |
| $\nu_e + e^- \to \nu_e + e^-$ | $32 G_F^2 [(C_A + C_V)^2 (p_1 \cdot p_2)(p_3 \cdot p_4) +$ |
| | $(C_A - C_V)^2 (p_1 \cdot p_4)(p_2 \cdot p_3) -$ |
| | $(C_V^2 - C_A^2) m_e^2 (p_1 \cdot p_3)]$ |
| $\nu_e + e^+ \to \nu_e + e^+$ | $32 G_F^2 [(C_A + C_V)^2 (p_1 \cdot p_4)(p_2 \cdot p_3) +$ |
| | $(C_A - C_V)^2 (p_1 \cdot p_2)(p_3 \cdot p_4) -$ |
| | $(C_V^2 - C_A^2) m_e^2 (p_1 \cdot p_3)]$ |
| $\nu_e + \overline{\nu}_e \to e^- + e^+$ | $32 G_F^2 [(C_A + C_V)^2 (p_1 \cdot p_4)(p_2 \cdot p_3) +$ |
| | $(C_A - C_V)^2 (p_1 \cdot p_3)(p_2 \cdot p_4) +$ |
| | $(C_V^2 - C_A^2) m_e^2 (p_1 \cdot p_2)]$ |
| $\nu_e + \overline{\nu}_e \to \nu_i + \overline{\nu}_i$ | $32 G_F^2 (p_1 \cdot p_4)(p_2 \cdot p_3)$ |

TABLE II. Possible $\mu, \tau$ neutrino processes and matrix elements. The definition of particle numbers is the same as in TABLE I.

| Process | $S \mid M \mid^2$ |
|---------|-------------------|
| $\nu_i + \overline{\nu}_i \rightarrow \nu_i + \overline{\nu}_i$ | $128 G_F^2 (p_1 \cdot p_4)(p_2 \cdot p_3)$ |
| $\nu_i + \overline{\nu}_j \rightarrow \nu_i + \overline{\nu}_j$ | $32 G_F^2 (p_1 \cdot p_4)(p_2 \cdot p_3)$ |
| $\nu_i + \nu_i \rightarrow \nu_i + \nu_i$ | $32 G_F^2 (p_1 \cdot p_2)(p_3 \cdot p_4)$ |
| $\nu_i + \nu_j \rightarrow \nu_i + \nu_j$ | $32 G_F^2 (p_1 \cdot p_2)(p_3 \cdot p_4)$ |
| $\nu_i + \overline{\nu}_e \rightarrow \nu_i + \overline{\nu}_e$ | $32 G_F^2 (p_1 \cdot p_4)(p_2 \cdot p_3)$ |
| $\nu_i + \nu_e \rightarrow \nu_i + \nu_e$ | $32 G_F^2 (p_1 \cdot p_2)(p_3 \cdot p_4)$ |
| $\nu_i + e^- \rightarrow \nu_i + e^-$ | $32 G_F^2 [(C_V + C_A - 2)^2 (p_1 \cdot p_2)(p_3 \cdot p_4) +$ |
| | $(C_A - C_V)^2 (p_1 \cdot p_4)(p_2 \cdot p_3) -$ |
| | $[(C_V - 1)^2 - (C_A - 1)^2] m_e^2 (p_1 \cdot p_3)]$ |
| $\nu_i + e^+ \rightarrow \nu_i + e^+$ | $32 G_F^2 [(C_V + C_A - 2)^2 (p_1 \cdot p_4)(p_2 \cdot p_3) +$ |
| | $(C_A - C_V)^2 (p_1 \cdot p_2)(p_3 \cdot p_4) -$ |
| | $[(C_V - 1)^2 - (C_A - 1)^2] m_e^2 (p_1 \cdot p_3)]$ |
| $\nu_i + \overline{\nu}_i \rightarrow e^- + e^+$ | $32 G_F^2 [(C_V + C_A - 2)^2 (p_1 \cdot p_4)(p_2 \cdot p_3) +$ |
| | $(C_A - C_V)^2 (p_1 \cdot p_3)(p_2 \cdot p_4) +$ |
| | $[(C_V - 1)^2 - (C_A - 1)^2] m_e^2 (p_1 \cdot p_2)]$ |
| $\nu_i + \overline{\nu}_i \rightarrow \nu_e + \overline{\nu}_e$ | $32 G_F^2 (p_1 \cdot p_4)(p_2 \cdot p_3)$ |
| $\nu_i + \overline{\nu}_i \rightarrow \nu_j + \overline{\nu}_j$ | $32 G_F^2 (p_1 \cdot p_4)(p_2 \cdot p_3)$ |